\documentstyle[sprocl,psfig]{article}

\def\Journal#1#2#3#4{{#1} {\bf #2}, #3 (#4)}

\def\NPB{{\em Nucl. Phys.} B}
\def\NPA{{\em Nucl. Phys.} A}
\def\PLB{{\em Phys. Lett.}  B}
\def\PRL{\em Phys. Rev. Lett.}
\def\PRD{{\em Phys. Rev.} D}
\def\ZPC{{\em Z. Phys.} C}
\def\PPNP{\em Progr. Part. Nucl. Phys.}

\def\be{\begin{equation}}
\def\ee{\end{equation}}
\def\bea{\begin{eqnarray}}
\def\eea{\end{eqnarray}}

\def\bq{\begin{eqnarray}}
\def\eq{\end{eqnarray}}
\newcommand{\bfk}{\mbox{\boldmath $k$}}
\newcommand{\bfy}{\mbox{\boldmath $y$}}
\newcommand{\pup}{p^\uparrow}
\newcommand{\pdown}{p^\downarrow}

\begin{document}

\title{THE POLARIZED NUCLEON IN QUARK MODELS}

\author{ALESSANDRO DRAGO}

\address{Dipartimento di Fisica, Universit\`a di Ferrara and INFN, Sezione
di Ferrara,\\ 
I-44100 FERRARA, ITALY\\E-mail: drago@fe.infn.it}


\maketitle\abstracts{A brief overview of various problems related to the 
description of a polarized proton in quark models is presented. Structure
functions are discussed both for longitudinal and transverse polarization.
A recently introduced quantity, relevant in the study of single-spin 
asymmetries, is shown to be in principle non-vanishing when 
computed in chiral models.}

\section{Introduction}

The possibility of describing a polarized nucleon in terms of effective degrees
of freedom is still an open problem. Various experiments have discovered
effects that cannot be explained within perturbative QCD, {\it e.g.}
the single-spin asymmetries found in proton-proton scattering.  
Even the so-called spin problem of the nucleon, namely the way in which
the total angular momentum of the system is distributed among the various
components, is not yet completely understood and substantially different
mechanisms have been proposed. 

In this contribution I will shortly review a few topics related to the
description of a polarized nucleon in high-energy experiments.
To this purpose, let me recall the expression of the three leading twist 
quark distribution functions:

unpolarized:

\be
q(x)=\frac{\sqrt{2}} {4\pi}\int{\rm d}\xi^-
e^{-i x p^+\xi^-}
\langle N|\psi_+ ^\dagger(\xi)
\psi_+ (0)| N\rangle
\vert_{\xi^+=\xi_\perp=0}
\ee

longitudinally polarized:

\be
\Delta q=\frac{\sqrt{2}} {4\pi}\int{\rm d}\xi^-
e^{-i x p^+\xi^-}
\langle N|\psi_+ ^\dagger(\xi)\gamma_5
\psi_+ (0)| N\rangle
\vert_{\xi^+=\xi_\perp=0}
\ee

transversely polarized:

\be
h_1(x)=\frac{\sqrt{2}} {4\pi}\int{\rm d}\xi^-
e^{-i x p^+\xi^-}
\langle NS_\perp|\psi_+ ^\dagger(\xi)
\gamma_\perp\gamma_5 \psi_+ (0)| NS_\perp\rangle
\vert_{\xi^+=\xi_\perp=0}.
\ee
The first two quantities are the well known distributions 
of the momentum and of
the helicity, respectively. The third one has been introduced 
recently \cite{jaffeji}
and is the distribution of the transversity in a transversely
polarized proton.

The previous quantities can be evaluated in any quark model which provides
the wave function of the nucleon.
What one obtains are the leading twist
contributions to the distributions evaluated at a very low
$Q_0^2$, the scale of the model. To compare with the experiments one
has to compute their evolution to larger value of $Q^2$ using DGLAP equations.

\section{Longitudinal polarization}

In Fig.1 results for the unpolarized and for the longitudinally polarized
structure functions are presented. They were obtained in Ref.[2],
using the so-called chiral chromodielectric model \cite{CDM}. The latter is
a non-topological soliton model in which chiral fields play a relatively
minor role. This is due to the absence in the model of solutions having 
non-zero winding number. As it appears, the computed longitudinally polarized
structure function largely overestimate the data, its first momentum being
$\Gamma_1^p=0.225$ (exp.$\sim$ 0.136). 

\begin{figure}[t]
\centerline{\hbox{
\psfig{figure=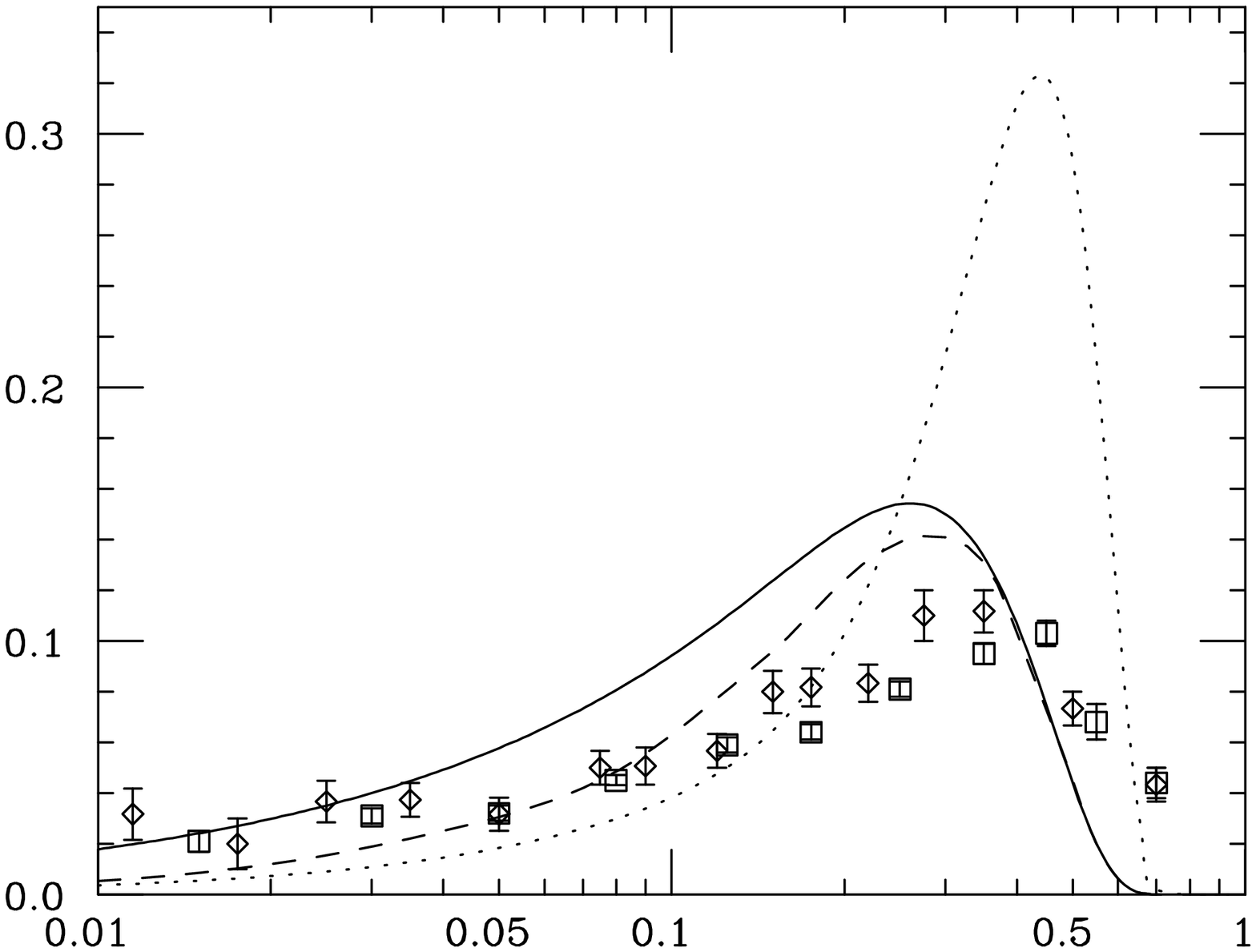,angle=90,height=7truecm,width=7truecm}
\psfig{figure=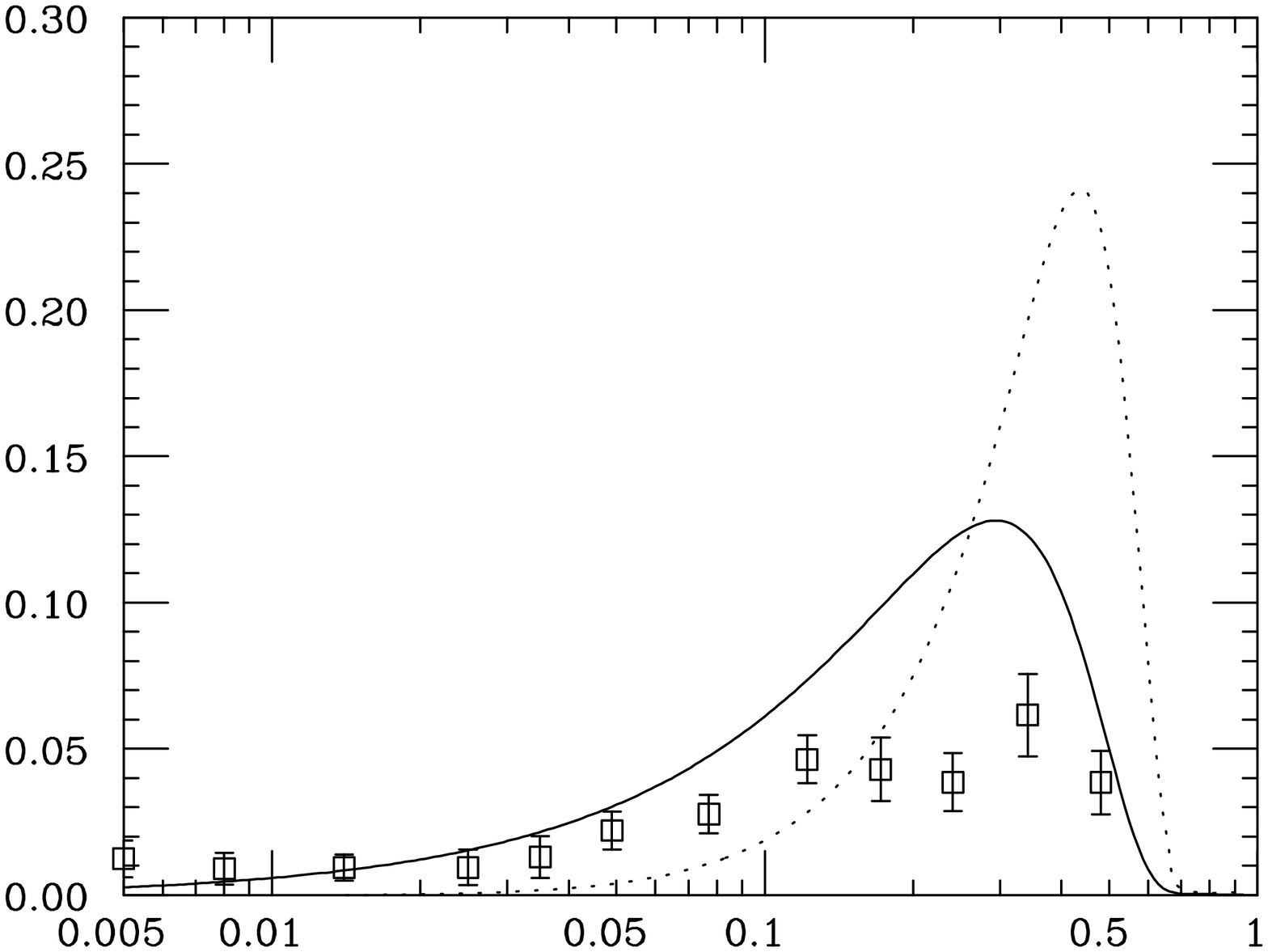,angle=90,height=7truecm,width=7truecm}
}}
\caption{Unpolarized non-singlet $F_2^p-F_2^n$ (left)
and polarized proton $x g_1^p$ (right) structure functions. The dotted curves
represent the leading-twist contributions at the model scale $Q_0^2$=0.16 
GeV$^2$. The solid curves are the evolved distributions at $Q^2$=4 GeV$^2$
(unpolarized) and at $Q^2$=10 GeV$^2$ (polarized), respectively. Diamonds
and square are the data. For details see Ref.[2].  }
\end{figure}

Chiral models in which the pion develops a non-trivial topology
are able to reduce the valence quark polarization, converting
spin into orbital angular momentum of the sea, 
represented by the chiral fields \cite{chiral}.
They offer wherefore a possible solution
of the spin problem. 

It must anyway be emphasized that the analysis of the DIS data
suggests a rather large polarization of the gluons \cite{altarelli}, 
possible only if
the latter are polarized already at very low $Q^2$. Moreover, QCD
sum rules results indicates that roughly half of the total angular
momentum is carried by the gluons \cite{sumrules}. 
In Ref.[7] the total angular
momentum carried by the gluons $J_g$ was estimated using the simple Isgur-Karl
model.  
At the scale of the model $Q_0^2=$0.25 GeV$^2$ roughly half of the
total angular momentum is attributed to the spin of the gluons 
$\Delta G\sim 0.24$. The
orbital angular momentum of the gluons turns out to be negligible at $Q_0^2$.
Performing a leading-order QCD evolution, $\alpha_s \Delta G$ is constant,
whereas $J_g$ increases slowly.  

At the moment it is impossible to decide which of the two solutions
of the spin problem is realized in nature. Probably the angular momentum
is carried partly by the sea and partly by the gluons. The forthcoming
experiments aiming to measure $\Delta G$ should clarify the situation.

\section{Transverse polarization}

The transversity distribution $h_1$ cannot be measured in fully
inclusive DIS, since it corresponds to a process in which 
the helicity of the struck quark is flipped. At the moment
no experimental data is available. $h_1$ could be measured in semi-inclusive
electron scattering or in transversely polarized Drell-Yan
processes.  

The QCD evolution of $h_1$ differs from that of $g_1$ since:
\begin{itemize}
\item
$h_1(x,Q^2)$ does not mix with the gluon distribution \cite{artru}
\item
its first momentum (tensor charge) decreases as 

$ \delta q(Q^2) = \delta q(Q_0^2) \, 
\left [ \frac{\alpha_s(Q_0^2)}{\alpha_s(Q^2)} 
\right ]^{-4/27}$
\item
at small x its splitting function $P_h(x)\sim 8x/3$ \cite{barone}, 
whereas $\Delta P_{qq}$
and $\Delta P_{qg}$ 
(the splitting functions for $g_1$)
behave as constant as $x\rightarrow 0$
\end{itemize}

\begin{figure}

\centerline{\hbox{
\psfig{figure=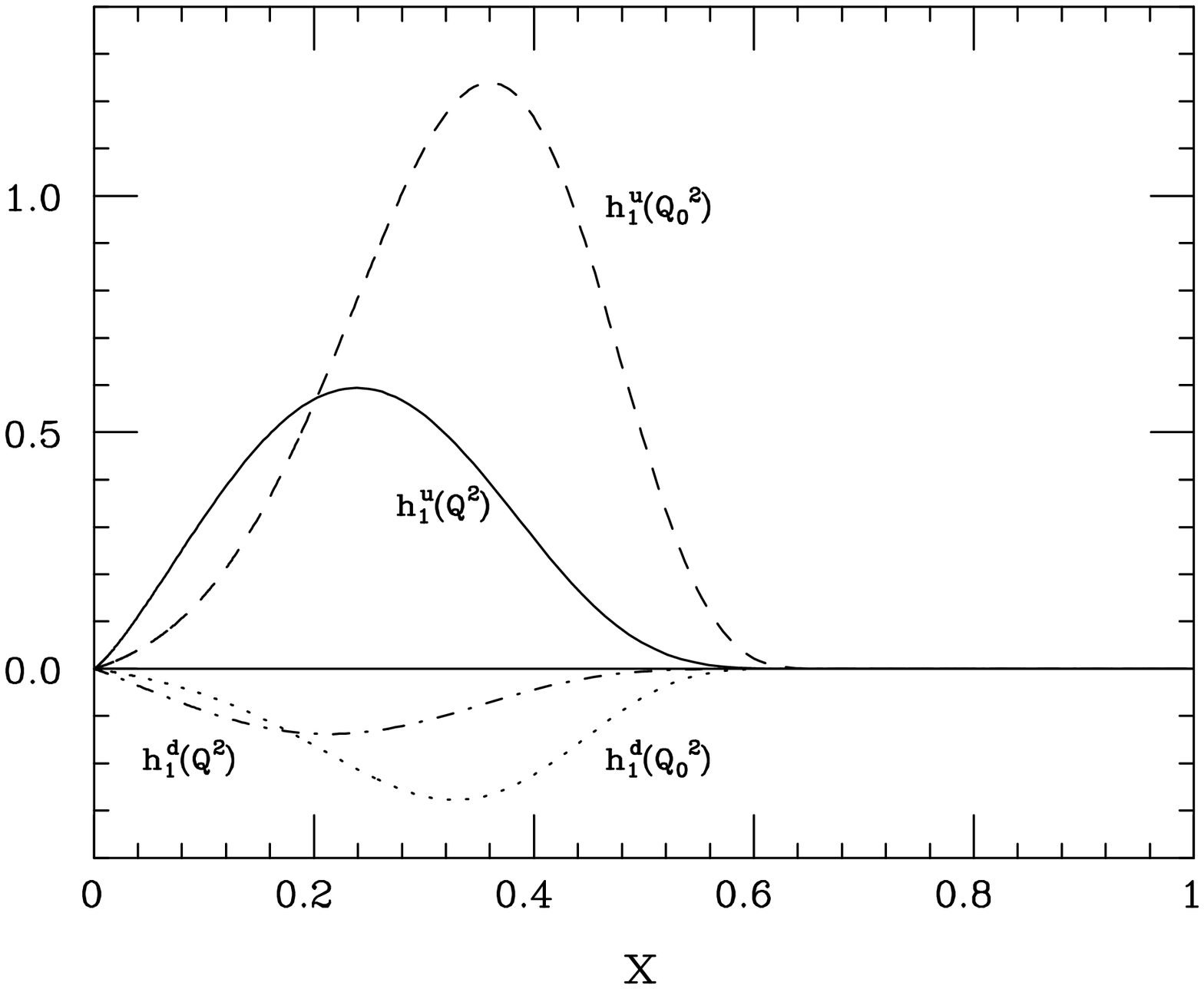,angle=90,height=7truecm,width=7truecm}
\psfig{figure=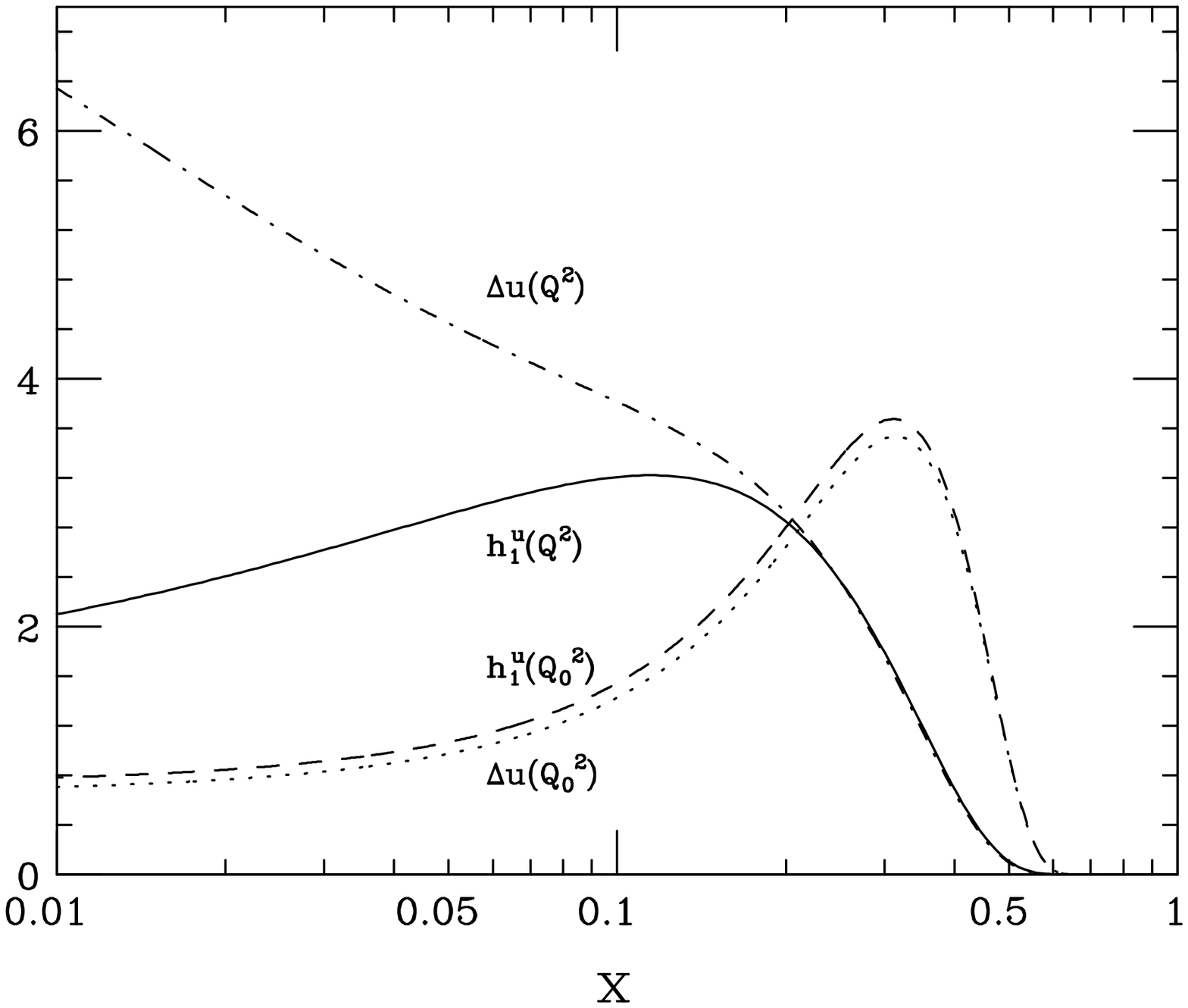,angle=90,height=7truecm,width=7truecm}
}}
\caption{Left: transversity distribution for $u$ and $d$ quarks
at the scale of the model and at $Q^2=$25 GeV$^2$. Right:
Helicity and transversity distribution for the $u$ quark. From Ref.[10].}
\end{figure}

In Fig. 2 estimates of $h_1$ made using the chiral chromodielectric model
are presented \cite{plb390}. As it can be seen, 
$h_1$ is not substantially different from $g_1$ at the scale of the model,
but due to the linear behaviour in $x$ of its splitting function
as $x\rightarrow 0$, $h_1$
is considerably smaller than $g_1$ at small $x$ after
being evolved to large $Q^2$. As a consequence, the 
double transverse Drell-Yan asymmetries turn out to be considerably
smaller than the longitudinal ones, making it very difficult
to measure $h_1$ through this process. The previous statements have been
confirmed by a model independent analysis \cite{prd}.

\section{Single spin asymmetries and chiral lagrangians}

Consider the plane defined by the momentum of the proton
and its {\it transverse} spin.
Recently a new quantity has been introduced  \cite{single}. 
It has the following partonic
interpretation: it gives the asymmetry
in the distribution of the momenta perpendicular to the above defined plane.
\be
\Delta^Nf_{a/p^\uparrow}(x_a,\bfk_{\perp a})
= \sum_{\lambda^{\,}_a}
\left[ \hat f_{a, \lambda^{\,}_a / p^\uparrow} (x_a,\bfk_{\perp a})
- \hat f_{a, \lambda^{\,}_a / p^\uparrow} (x_a, -\bfk_{\perp a}) \right] 
\ee

This quantity can be regarded as a single spin asymmetry 
for the $p^\uparrow \to a + X$ process and was introduced to describe
the process $p p^\uparrow\rightarrow\pi X$.
It can be written in terms of matrix elements of quark operators as:
\be
\Im{\rm m}\int{{\rm d}y^- {\rm d} \bfy_\perp\over(2 \pi)^3}
e^{-i x p^+ y^- + i \bfk_\perp\cdot\bfy_\perp}
\langle p, -|\bar\psi_a(0,y^-,y_\perp)
{\gamma^+\over 2}\psi_a(0)|p, + \rangle 
\ee

It has been shown by Collins \cite{collins}
that, {\it if the flavor $a$ is not touched by
time reversal}, the previous quantity vanishes due to time reversal invariance
of QCD. Indeed,
no time-reversal even observable can be 
constructed with two independent momenta and one spin vector.

In Weinberg's book on field theory \cite{weinberg} it is shown that
a field can transform under time reversal in a more complicated
way than just getting a phase.  In particular
a doublet of fields can mix under time reversal.
Now, in chiral lagrangians the flavor is not a dummy index.
Let me consider the single-quark Dirac equation in the presence of static
chiral fields:
$$
[k_\mu \gamma^\mu -g(\sigma+i\gamma_5\vec\tau\cdot\vec\pi)]u(k)=0. 
$$
Seeking the time reversed solution of the same equation
($\bfk\rightarrow -\bfk$) we get:
$$[k_\mu \gamma^\mu -g(\sigma-i\gamma_5\vec\tau\,^T \cdot\vec\pi)]
\gamma_5 C u^*(\tilde k)=0 
$$
where $\tilde k=(k_0,-\bfk)$ and $C=i\gamma_0 \gamma_2$.  
The term containing
the pion has been modified by the previous transformation.  
Since
$(-i\tau_2)(-\vec\tau\,^T)(i\tau_2)=\vec\tau$, the time reversed
solution reads $(-i\tau_2)\gamma_5 C u^*(\tilde k)$. This
means that
quark states of fixed isospin are in general not eigenstates of the
chiral hamiltonian.
For instance, taking the hedgehog form for the pion field,
$\vec\pi=\hat r \phi(r)$, the spin-isospin wave function is given by
$|h\rangle=[\,|u + \rangle-|d - \rangle]/\sqrt 2$.

If time reversal mixes up and down quarks the asymmetry function
$\Delta^Nf_{a/\pup}$ need not be zero and single spin
asymmetries in inclusive DIS are allowed \cite{hep}:

$$
\frac{d\sigma^{\ell\pup\to\ell X}}{dx\,dQ^2}-
\frac{d\sigma^{\ell\pdown\to\ell X}}{dx\,dQ^2} 
= \sum_{q}\int d\bfk_{\perp}\,
\Delta^{N}f_{q/\pup}(x,\bfk_{\perp})\,
\frac{d\hat{\sigma}^{\ell q\to\ell q}}{dQ^2}(x,\bfk_{\perp})\nonumber
$$
It would be extremely interesting to test this asymmetry with an experiment. 

\section{Conclusions}

\begin{itemize}
\item
Longitudinal Polarization
\begin{itemize}
\item
Chiral fields alone could be not sufficient to solve the {\it spin problem}
\item
The spin carried by polarized gluons can be computed in quark models
\item
The gluon contribution to the spin 
turns out to be of the right order of magnitude and of the right sign
\end{itemize}
\item
Transverse Polarization
\begin{itemize}
\item
The transverse distribution $h_1(x)$ evolves differently from 
the longitudinal one
\item
At small $x$ $h_1(x)$ is considerably smaller than $g_1(x)$
\item
It is therefore difficult to measure  $h_1(x)$
\end{itemize}
\item
Single-spin asymmetries
\begin{itemize}
\item
They are not allowed in DIS if flavor is untouched by time reversal
\item
They are possible in chiral models and could be tested experimentally.

\end{itemize}
\end{itemize}

\section*{Acknowledgments}
The material presented in this contribution 
is based on works done in collaboration 
with M. Anselmino, T. Calarco, M. Fiolhais, F. Murgia and in particular
V. Barone with whom I discussed 
all the problems and the possibilities considered here. L. Caneschi
helped with stimulating criticism.

\end{document}